\documentclass[preprint,aps,prl,superscriptaddress]{revtex4-1}

\usepackage{graphicx}  
\usepackage{dcolumn}   
\usepackage{bm}        
\usepackage{amssymb}   
\usepackage{amsmath}   
\usepackage{mathtools} 
\usepackage{tabularx}  
\usepackage{lipsum}    
\graphicspath{{figs/}} 

\begin{document}
\title{\bf{Chiral self-sorting of active semiflexible filaments with intrinsic curvature}} 
\author{Jeffrey M. Moore}
\author{Matthew A. Glaser} 


\affiliation{Department of Physics, University of Colorado, Boulder, CO 80309}

\author{Meredith D. Betterton}
\affiliation{Department of Physics, University of Colorado, Boulder, CO 80309}
\affiliation{Department of Molecular, Cellular, and Developmental Biology, University of Colorado, Boulder, CO 80309}

\date{\today}

\begin{abstract}

    Many-body interactions in systems of active matter can cause particles to move collectively and self-organize into dynamic structures with long-range order. In cells, the self-assembly of cytoskeletal filaments is critical for cellular motility, structure, intracellular transport, and division. Semiflexible cytoskeletal filaments driven by polymerization or motor-protein interactions on a two-dimensional substrate, such as the cell cortex, can induce filament bending and curvature leading to interesting collective behavior. For example, the bacterial cell-division filament FtsZ is known to have intrinsic curvature that causes it to self-organize into rings and vortices, and recent experiments reconstituting the collective motion of microtubules driven by motor proteins on a surface have observed chiral symmetry breaking of the collective behavior due to motor-induced curvature of the filaments. Previous work on the self-organization of driven filament systems have not studied the effects of curvature and filament structure on collective behavior. In this work, we present Brownian dynamics simulation results of driven semiflexible filaments with intrinsic curvature and investigate how the interplay between filament rigidity and radius of curvature can tune the self-organization behavior in homochiral systems and heterochiral mixtures. We find a curvature-induced phase transition from polar flocks to self-sorted chiral clusters, which is modified by filament flexibility. This phase transition changes filament transport from ballistic to diffusive at long timescales.
\end{abstract}

\maketitle




Active matter systems with particles that are driven can self-organize into structures that exhibit emergent dynamical order on length scales many times greater than the scale of a single particle. In cells, the dynamic self-assembly of cytoskeletal filaments driven by polymerization, motor proteins, and crosslinkers is necessary for the production of macromolecular assemblies, such as the mitotic spindle~\cite{hyman96, nedelec97, surrey01, needleman17}. Most previous theoretical studies of cytoskeletal self-assembly have assumed rigid and straight filaments, but recent experiments have identified self-organization behavior due to filament curvature. Experiments reconstituted collective self-assembly by driving filaments on a surface with bound motor proteins~\cite{butt10, schaller10, schaller11, liu11, sumino12, huber18, kim18}. Microtubules can undergo chiral symmetry breaking due to induced curvature mediated by motors, causing the filaments to self-organize into a dynamic lattice of vortices~\cite{sumino12},  or rotating polar streams~\cite{kim18}. In addition, \emph{in vivo} observations of the bacterial cytoskeletal filament FtsZ found that intrinsic curvature led to the formation of dynamic bundles of curved rings and vortices~\cite{meier14, loose14}. These findings motivate further study of the collective behavior of driven filaments with curvature.


Previous theoretical studies of active systems with chirality have also taken their inspiration from biological systems. Swimming sperm cells~\cite{riedel05, friedrich07}, confined bacteria~\cite{wensink12b, lowen16}, and \emph{E. coli} swimming near a solid interface~\cite{diluzio05, lauga06} all exhibit chiral behavior. Brownian dynamics simulations have been used to predict the collective behavior of chiral microswimmers, and found phase separation of particles into spatially-segregated clusters~\cite{liebchen17a, liao18, levis18, levis19}, activity-induced chiral self-sorting of heterochiral mixtures~\cite{nguyen14, levis19a}, demixing of differently-shaped particles~\cite{wensink14}, and stable vortex arrays of circle swimmers~\cite{kaiser13a, yang14}. There have, however, been few theoretical studies of active curved filaments. Previous work includes simulations of FtsZ filaments with varying particle density and noise strength~\cite{denk16}, and passive filaments in the presence of a swirling flow~\cite{kuchler16}. FtsZ, microtubules, and f-actin each have different stiffness and intrinsic or induced curvature. Therefore, this work focuses on self-organization of driven curved filaments over a range of bending rigidity and radius of curvature. 

In this work, we present results of Brownian dynamics simulations of active semiflexible filaments. We study both homochiral and heterochiral systems and vary filament bending rigidity, radius of curvature, packing fraction, and aspect ratio. The collective behavior depends strongly on filament flexibility and curvature: as the curvature increases, there is a dramatic transition of the phase behavior from polar flocking to self-sorted chiral clusters. However, the transition is weakened or lost if the filaments are too flexible.

\begin{figure}[tb] \centering
  \includegraphics[width=0.75\textwidth]{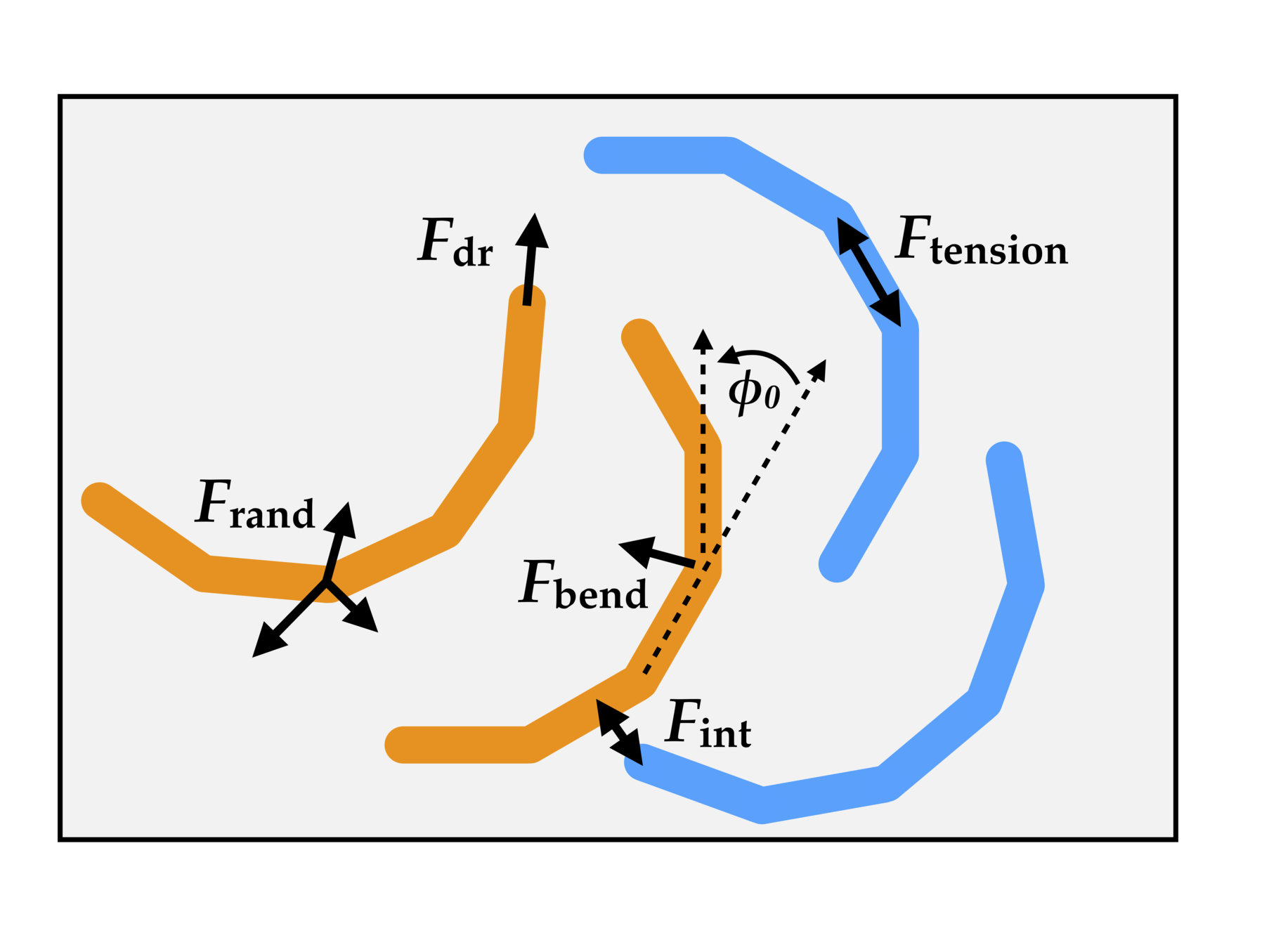}
  \caption{\small Schematic of the model and forces acting on filaments. Filaments are discretized wormlike chains composed of multiple rigid segments. Filaments are subject to driving forces $\mathbf{F}_{\text{dr}}$, a bending force $\mathbf{F}_{\text{bend}}$, interaction forces $\mathbf{F}_{\text{int}}$, a tension force $\mathbf{F}_{\text{tension}}$, and random forces $\mathbf{F}_{\text{rand}}$. The polar driving forces are tangent to all filament segments and control system activity. The bending force maintains an intrinsic curvature with equilibrium angle $\phi_0$ between filament segments. The interaction forces model steric repulsion between filaments. The tension force maintains the rigid segment length constraints. The random thermal forces give rise to Brownian motion of the filament.}
  \label{fig:model}
\end{figure}



Our Brownian dynamics simulations are two-dimensional with periodic boundary conditions of $N$ filaments with intrinsic curvature. Following previous work~\cite{moore20, moore20a}, the filaments in our simulations are modeled as discretized wormlike chains~\cite{kratky49} composed of $n$ sites and $n-1$ rigid segments of fixed length $a$. We adopt the algorithm of Montesi et. al~\cite{montesi05} for constrained Brownian dynamics of bead-rod wormlike chains. The $n$ site positions are updated according to inertialess Langevin equation
\begin{equation}
    \dot{\mathbf{r}}_{i} = \boldsymbol{\zeta}_{i}^{-1}\mathbf{F}_{i},
\end{equation}
where $\boldsymbol{\zeta}$ is an anisotropic friction tensor with parallel and perpendicular components $\zeta_{\perp} = 2 \zeta_{\parallel}$, and site forces
\begin{equation}
    \mathbf{F}_i = \mathbf{F}_i^{\text{dr}} + \mathbf{F}_i^{\text{bend}} + \mathbf{F}_i^{\text{int}} + \mathbf{F}_i^{\text{tension}} + \mathbf{F}_i^{\text{rand}},
\end{equation}
include driving forces $\mathbf{F}_{\text{dr}}$, bending forces $\mathbf{F}_{\text{bend}}$, filament-filament interaction forces $\mathbf{F}_{\text{int}}$, tension forces $\mathbf{F}_{\text{tension}}$, and random thermal forces $\mathbf{F}_{\text{rand}}$ (Fig.~\ref{fig:model}). 

Filament activity is induced by a polar driving force per unit length $f_{\text{dr}}$ acting tangent to each of the filament segments, $\mathbf{F}_i^{\text{dr}} = a f_{\text{dr}} \mathbf{u}_i$. The activity of the system is measured by the P\'eclet number, the ratio of active to diffusive transport. For straight and rigid filaments, the characteristic timescale for active transport is the time required for the filament moving at velocity $v$ to traverse a distance equal its own length $L$, $\tau_A = L/v = \zeta_{\parallel}L/F_{\text{dr}}$. The time for a filament to diffuse its own length is $\tau_D = L^2/D_{\parallel} = \zeta_{\parallel} L^2/k_BT$; thus the P\'eclet number is $\text{Pe} = \tau_D/\tau_A = f_{\text{dr}}L^2/k_BT$. In the case of curved filaments, the mean-squared displacement and therefore the effective active transport can differ, as discussed later. We continue to use this definition of the P\'eclet number for consistency.

Filament bending forces are derived from the bending potential of a continuous wormlike chain $\frac{\kappa}{2}\int_{0}^{L}(R^{-1}(s) - R_0^{-1})^2\,ds$, where $R(s)$ is the local radius of curvature with respect to the contour length $s$ for a length $L$ filament, and $R_0$ is a radius of curvature that corresponds to an intrinsic curvature per unit length $R_0^{-1} = d\phi/ds$. The persistence length of the filament is related to the bending rigidity as $L_pk_BT = \kappa$. In a discrete wormlike chain model, the bending potential is
\begin{equation}\label{eqn:ubendic}
  U_{\text{bend}} = - \frac{\kappa}{a}\sum_{i=2}^{n-1} \cos{(\theta_{i,i-1} - \phi_0)},
\end{equation}
where $\theta_{i, j} = \arccos{(\mathbf{u}_i\cdot\mathbf{u}_{j})}$ is the angle between segments $i$ and $j$, and $\phi_0=ad\phi/ds$ corresponds to the expected angle between two segments of length $a$ and intrinsic curvature per unit length $d\phi/ds$. 

Steric repulsion between filaments $\mathbf{F}_{\text{int}}$ is modeled by a Weeks-Chandler-Andersen (WCA) potential, 
\begin{equation}
    U_{\text{WCA}} = \Big( 4 \epsilon \big( (\frac{\sigma}{r_{ij}})^{12} - (\frac{\sigma}{r_{ij}})^6 \big) + \epsilon \Big) \Big(1 - \Theta(2^{1/6}\sigma)\Big),
\end{equation}
where $r_{ij}$ is the minimum distance between two filaments, $\sigma$ is the filament diameter, $\epsilon=k_BT$ is the energy scale and $\Theta$ is the Heaviside step function.

To model the wormlike chain as an inextensibile filament with rigid bonds, the model is subject to constraints for the segment length $|\mathbf{r}_{i} -\mathbf{r}_{i-1}| = a$. A tension force is necessary to prevent external forces from violating the constraints, restricting the dynamics of the $N$ filament sites to a constrained subspace~\cite{morse04}.

The random forces $\mathbf{F}_{\text{rand}}$ applied to each site are due to thermal contact with a heat bath at temperature $T$. The components of the random forces $F^{(k)}_{\text{rand}}$ are delta-correlated random variables with mean $\langle F^{(k)}_{\text{rand}}(t) \rangle = 0$ and variance $\langle F^{(k)}_{\text{rand}}(t)F^{(k)}_{\text{rand}}(t') \rangle = 2 k_B T \zeta \delta(t-t')$ as dictated by the fluctuation-dissipation theorem. The random forces are then geometrically projected to prevent violation of the segment length constraints (see ESI).

The characteristic length and energy scales in our simulation are the filament diameter $\sigma$ and $k_BT$.  We use dimensionless units: the dimensionless lengths for our system are the aspect ratio $\tilde L = L/\sigma$, the bending rigidity $\tilde \kappa = \kappa /Lk_BT = L_p/L$, and radius of curvature $\tilde R = R/L$. The packing fraction $\phi = A_{\text{fil}}/A_{\text{sys}}$, where $A_{\text{fil}}$ is the combined area of all $N$ filaments and $A_{\text{sys}}$ is the periodic area of the system. In our simulations, we vary the  bending rigidity from $\tilde \kappa = 10$--$1000$ and the radius of curvature from $\tilde R = 0.25$--$2$, as well as straight filaments with $\tilde R = \infty$. We also examine two filament densities $\phi=0.25$ and $0.5$, two aspect ratios $\tilde L=10$ and $20$, and both homochiral systems and 1:1 heterochiral mixtures. Simulations were run for $10^3-10^4 \tau_A$. Further discussion of computational details is available in the ESI.

\begin{figure}[tb] \centering
  \includegraphics[width=\textwidth]{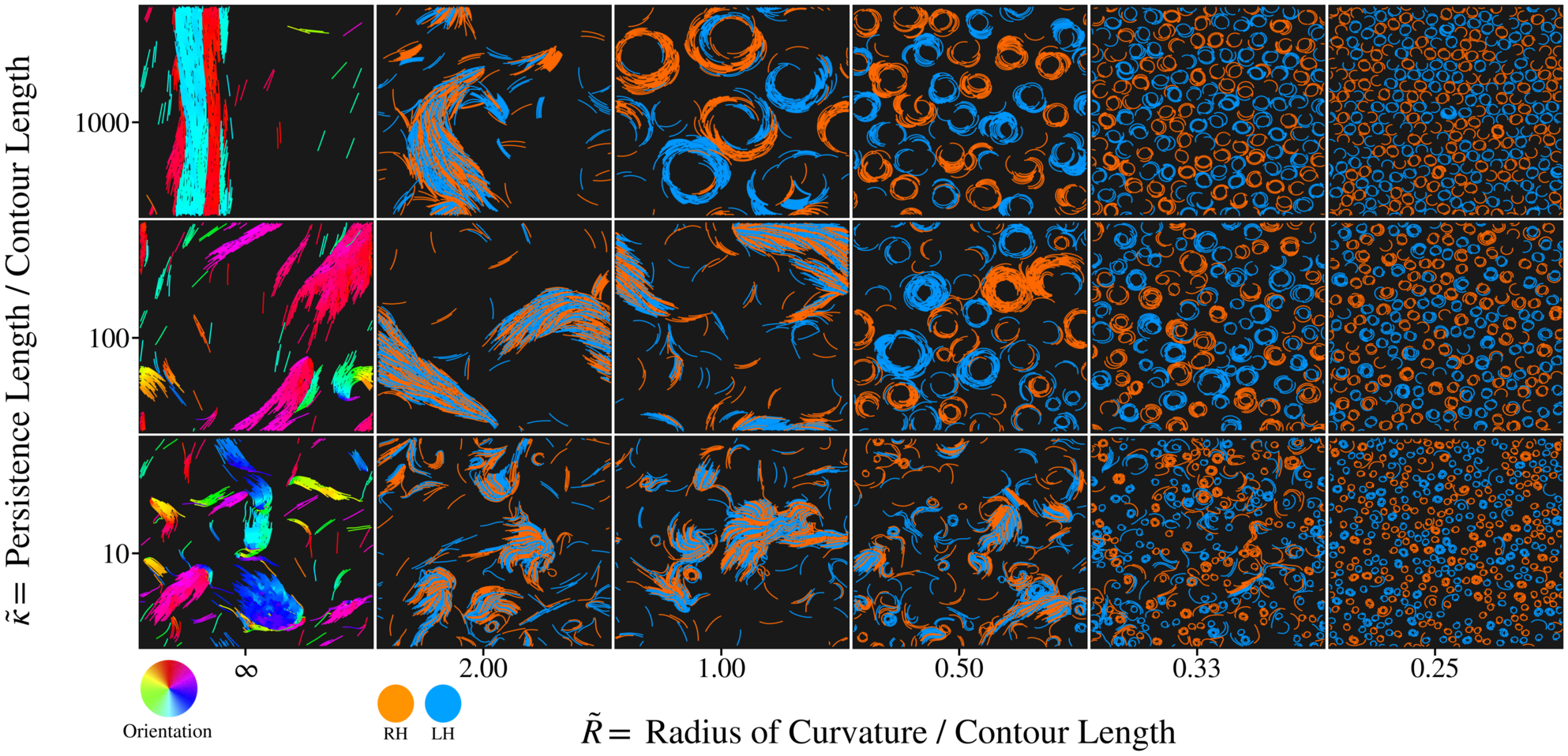}
  \caption{\small Simulation images depicting the collective behavior for heterochiral filaments of $\tilde L = 20$ and $\phi = 0.25$ for different values of filament stiffness $\tilde \kappa$ and radius of curvature $\tilde R$. The filaments are colored according to their orientation in the case of $\tilde R = \infty$ and according to their handedness otherwise, with orange and blue labeling curved filaments with right- and left-handed chirality respectively.}
  \label{fig:diagram}
\end{figure}

Driven curved filaments undergo a dramatic phase transition from polar flocks to chiral clusters, induced by filament curvature and tuned by filament flexibility (see Fig.~\ref{fig:diagram}). Driven achiral filaments form polar flocks or a swirling phase depending on filament rigidity and density. Chiral filaments with large radius of curvature form flocks of bent heterochiral filaments, leading to curved flock trajectories and buckling at higher densities. At low radius of curvature, there is a transition from polar heterochiral flocks to self-sorted homochiral clusters. Filament flexibility delays the onset of chiral clustering as the radius of curvature decreases.


In the absence of curvature, achiral driven filaments aggregate into well-known polar flocks~\cite{vicsek95, toner95, gregoire04, ginelli10, kuan15, ginelli16, moore20a}, with structural order that is dependent on filament bending rigidity, as shown in previous work~\cite{duman18, moore20a}. Driven flexible filaments can form an active disordered phase of swirling defects~\cite{lowen16, sanchez12, opathalage19}. Filaments with large radius of curvature ($\tilde R = 2$) can form heterochiral flocks composed of filaments with mixed handedness that are stable due to the deformability of the filaments.

With decreasing radius of curvature, the system undergoes a curvature-induced phase transition from polar flocks to chiral clusters. The lower radius of curvature increases the bending energy required to deform filaments into stable heterochiral flocks, so activity drives filaments into spatially-segregated domains of high filament density. Filaments of opposite handedness are rarely colocalized due to frequent collisions and fast decorrelation of their trajectories. Chiral self-sorting has been reported in previous studies of active chiral systems with spinning rotors~\cite{nguyen14, scholz18} and Vicsek-type models of chiral microswimmers~\cite{levis19a}, where self-sorting occurs at a macroscopic scale. In our model, steric repulsion between filaments and curved filament shape cause filaments to self-organize into a disordered lattice of self-sorted chiral vortices. Filament flexibility delays the onset of the phase transition to chiral clusters because it allows greater deformation away from the preferred curvature. Filament bending rigidity therefore tunes the long-range structural order of driven achiral filaments and the self-sorting behavior of active curved filaments.

In the flocking phase, rigid filaments form giant flocking domains that dominate the structure of the system, leading to strong nematic alignment. Flexible filaments form coexisting flocks with a distribution of sizes. The flocks are dynamic structures, with filaments continuously joining and leaving the flocks, though giant flocks are more stable and long-lived due to a large number of kinetically-trapped filaments at the flock interior~\cite{liu98, gravish12, kuan15, avendano17}. When the radius of curvature is large, curved filaments form heterochiral flocks due to the finite rigidity of the filaments. Comoving filaments with opposite chirality deform into internally jammed polar flocks. Filaments in the flock interior have low curvature, and can maintain stable giant flocks and nematic bands. However, more rigid filaments ($\tilde \kappa=1000$) tend to form packed layers reminiscent of smectic liquid crystals~\cite{leube90, brand92, avendano17}. This packing reduces interweaving between filament layers, and thereby decreases the stability of large flocks, even in entirely homochiral systems (see ESI). Therefore, filament curvature causes rigid filaments to have lower long-range structural order than flexible filaments, a reversal of what has been previously reported for achiral semiflexible filament behavior~\cite{duman18, moore20a}.

To better understand how curvature and bending rigidity affect the self-organization of the chiral cluster phase, we quantified the structure and sorting in our systems by measuring the radial distribution function. The radial distribution function $g(r)$ is defined for a 2D system with particle density $\rho$ such that $\langle 2 \pi r dr \rangle \rho g(r)$ is the number of particles in a circular annulus of width $dr$ at a distance $r$ from a reference particle, averaged over the particle ensemble: 
\begin{equation} \label{eqn:rdf}
    g(r) = \frac{2 \pi}{N\rho} \sum_{i=1}^N \int_{r}^{r+dr'}\sum_{j \neq i} \delta \big(\mathbf{r}' - (\mathbf{r_j} - \mathbf{r_i})\big)r'\,dr',
\end{equation}
where $\delta(x)$ is the Dirac delta function. The radial distribution functions for homochiral and heterochiral particles, $g_+(r)$ and $g_-(r)$, were also calculated by adding the factor $\delta_{\pm\chi_i \chi_j}$ in the second sum in Eqn.~\ref{eqn:rdf}, where $\delta_{ij}$ is the Kronecker delta function and $\chi = \pm 1$ for right- and left-handed chiral particles, respectively.

\begin{figure}[tb] \centering
  \includegraphics[width=\textwidth]{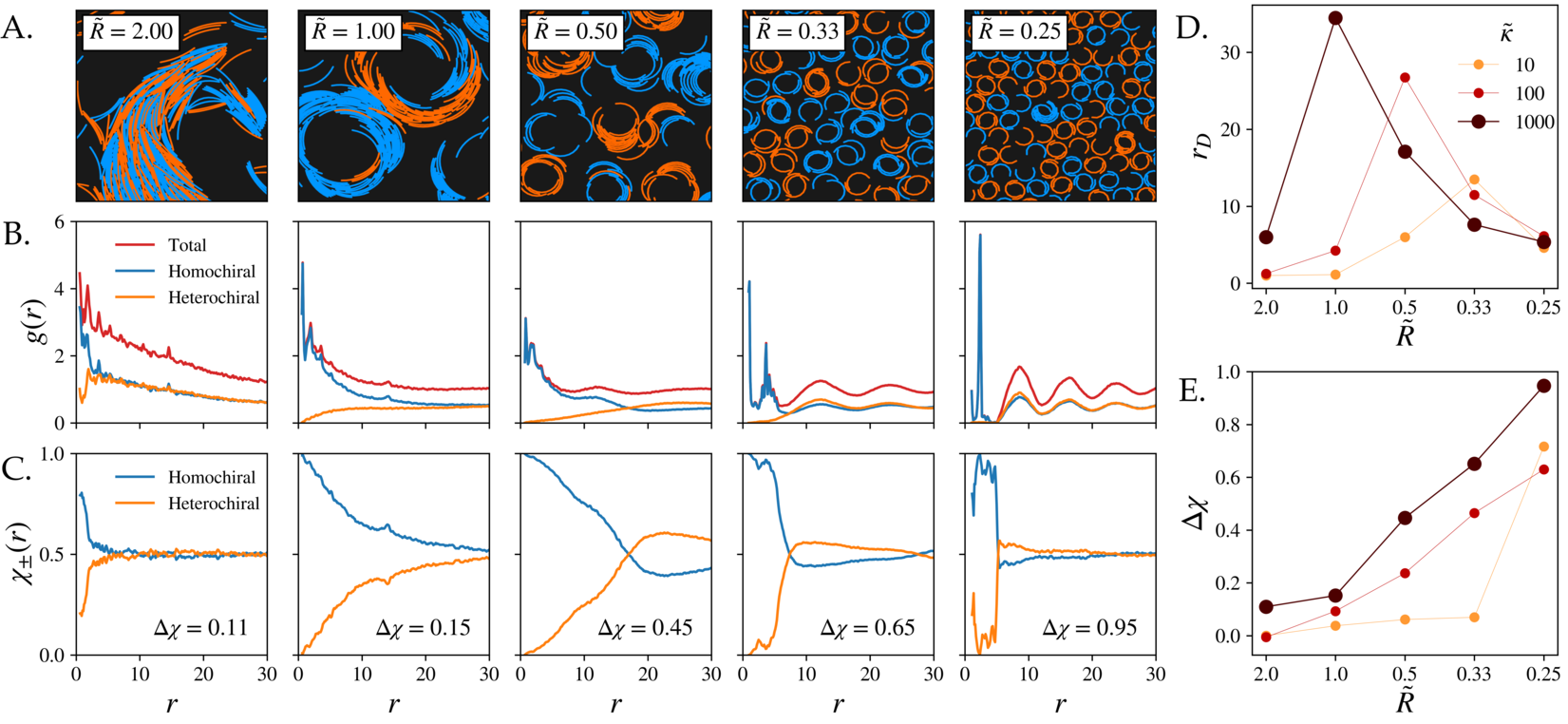}
  \caption{\small A) Simulation images of collective behavior for different filament radius of curvature, with $\tilde L = 20$, $\phi=0.25$, and $\tilde \kappa = 1000$. B) Radial distribution function $g(r)$ for filaments corresponding to the simulation images in A. The distribution function is plotted as the total distribution between all filaments (red), between homochiral filaments (blue), and between heterochiral filaments (orange). C) Chiral order parameter $\chi_{\pm}(r)$ for homochiral (blue) and heterochiral (orange) filaments. The corresponding chiral sorting parameter $\Delta \chi$ is included in the bottom right of the figures. D) Domain radius $r_D$ versus $\tilde R$ plotted for $\tilde L=20$, $\phi=0.25$. E) Chiral sorting parameter $\Delta \chi$ versus $\tilde R$ for $\tilde L=20$, $\phi=0.25$.} 
  \label{fig:sorting}
\end{figure}

Activity drives an increase in local particle density due to flocking and clustering behavior, indicated by a peak in $g(r)$ for $r\approx\sigma$ (Fig.~\ref{fig:sorting}B). In the chiral cluster phase, the peak is largely composed of homochiral filaments due to chiral sorting. The local density of heterochiral particles is suppressed below a length scale that specifies the sorted domain size, which depends on the radius of curvature and flexibility. 

\begin{figure}[tb] \centering
  \includegraphics[width=\textwidth]{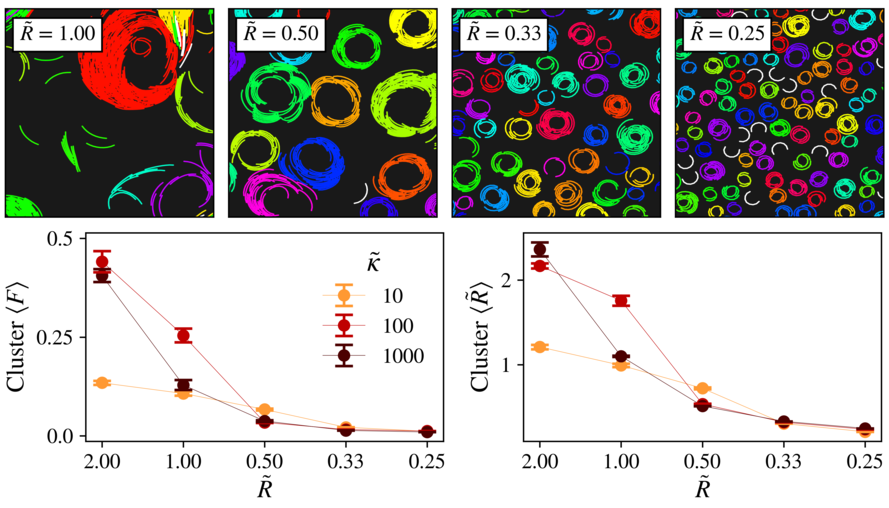}
  \caption{\small Top: simulation images of filament clusters for different filament radius of curvature. Clusters are labeled by color. Unclustered filaments are shown in white. Bottom left: time-averaged cluster filament fraction $\langle F \rangle$, which is the average number of filaments per cluster normalized by the number of filaments in the simulation, and plotted as a function of filament radius of curvature. Bottom right: mean cluster radius of curvature $\langle \tilde R \rangle$ as a function of filament radius of curvature $\tilde R$.}
  \label{fig:cluster}
\end{figure}

To quantify chiral sorting, we define a chiral structural order parameter $\chi_{\pm}(r) = g_{\pm}(r)/g(r)$ that measures the average homo/heterochiral sorting as function of distance $r$ from a reference particle. For a sorted system, $\chi_+(r)$ is strictly greater than $\chi_-(r)$ for $r < r_{\text{D}}$, where $r_{\text{D}}$ is the sorted domain radius, defined to be the first point at which $\chi_+(r) = \chi_-(r)$. The extent of chiral sorting within the domain is given by the chiral sorting order parameter
\begin{equation}
    \Delta \chi = \frac{\int_0^{r_D} \big(g_+(r) - g_-(r)\big) r\,dr}{\int_0^{r_D} g(r) r\,dr},
\end{equation}
which is the fraction of homochiral particles within the sorted domain, and varies from $0$--$1$ due to the definition of $r_D$. Thus $r_D$ measures the typical size of sorted domains and $\Delta \chi$ the degree of sorting within a domain.

The domain radius $r_D$ is largest near the transition between heterochiral flocking behavior and chiral self-sorting due to the large $\tilde R$ of filaments within the sorted cluster. $r_D$ typically decreases as $\tilde R$ decreases, as one would expect due to filament trajectories following smaller circular paths (Fig.~\ref{fig:sorting}D). Interestingly, the degree of chiral sorting increases with decreasing $\tilde R$ (Fig.~\ref{fig:sorting}E). This is due to smaller domains having more efficient packing within the system, and therefore fewer collisions between clusters result in fewer opportunities for filaments to intermix between sorted domains. Filament flexibility weakens chiral sorting due to filament deformation, consistent with our observations that flexibility weakens structural order in systems of driven filaments~\cite{moore20a}. Flexible filaments collide and form heterochiral flocks, which perturb chiral clusters with collisions, increasing mixing. In addition, the transition between phases of heterochiral flocking and chiral clusters is sharper for rigid filaments, whereas flexible filaments exhibit coexistence between the two states, and only become sorted at the smallest radius of curvature examined here.
.

To better understand the dynamical behavior of filament clusters, we developed a method to identify clustered filaments. Filaments were clustered by their centers of curvature $\mathbf{r}_c(t)$, determined from the filaments' instantaneous radius of curvature $R(t)$ averaged over the contour length of the filament (Fig.~\ref{fig:cluster}). Cluster positions are the average of their constituent filament centers of curvature, $\mathbf{r}_C(t) = \frac{1}{n}\sum_{i}^n \mathbf{r}_c^{(i)}(t)$, and the cluster radius is defined to be the average of the constituent filament radius of curvature $R_c(t) = \frac{1}{n}\sum_i^n R_i(t)$. Unclustered filaments can join an existing cluster when $|\mathbf{r}_c^{(i)}(t) - \mathbf{r}_C(t)| \leq R_c(t)$ for a time interval of $\tau_A$ (see ESI).

The mean cluster radius $\langle \tilde R \rangle$ decreases with radius of curvature as expected. The mean number of filaments in an average cluster also decreases with cluster size, since steric effects put an upper limit on the filament density within small clustered domains. Filament number is reported as a fraction of filaments in the system per cluster on average $\langle F \rangle$, with brackets $\langle ...\rangle$ denoting a time and ensemble average. Only clusters with lifetimes longer than $\tau_A$ are considered in our analysis to ensure a low false discovery rate. 



\begin{figure}[tb] \centering
  \includegraphics[width=\textwidth]{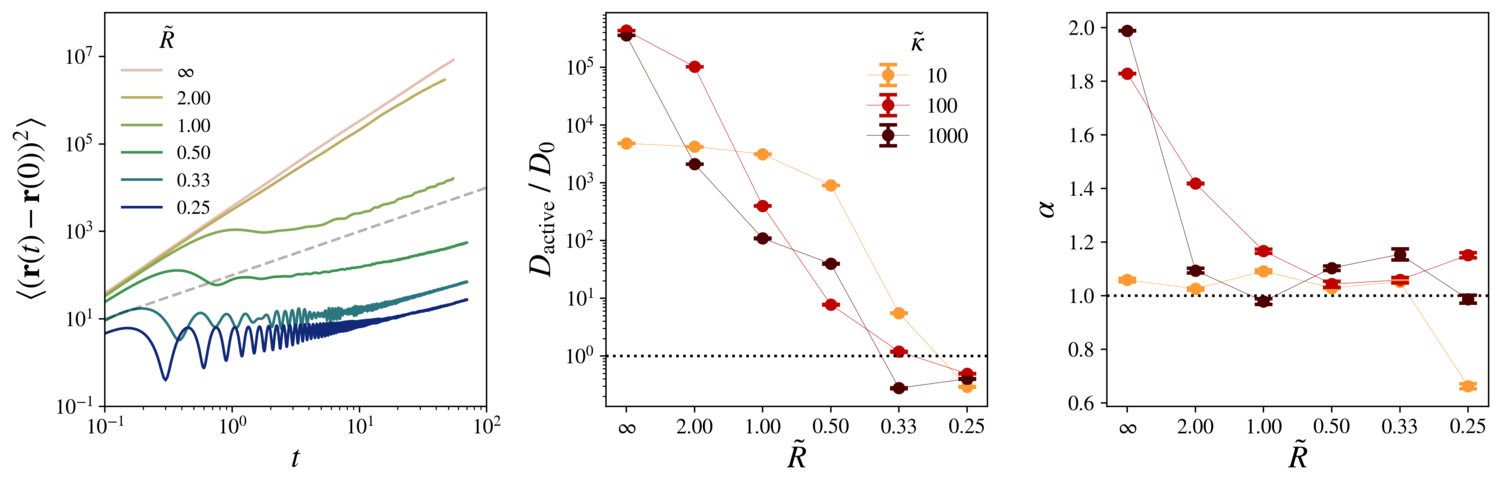}
  \caption{\small Left: mean-squared displacement (MSD) for active filaments of length $\tilde L=10$, filament density $\phi=0.25$, and stiffness $\tilde \kappa=1000$ for different values of radius of curvature. The average MSD is subdiffusive at short timescales and becomes diffusive at longer timescales. Center: Effective diffusion constants $D_{\text{active}}$ extracted from the final $20\tau_A$ of accessible simulation time assuming linear time scaling. The effective diffusion constant is normalized by diffusion constant of inactive filaments $D_0$ for otherwise identical simulation parameters (see ESI). Right: time scaling exponent $\alpha$ extracted from the late-time MSD used to calculated $D_{\text{active}}$. The transport of filaments that form smaller flocks and clusters approach diffusive transport at late times, while the ballistic motion of giant flocks produces superdiffusive behavior ($\alpha>1$).}
  \label{fig:msd}
\end{figure}

The circular motion of clustered filaments drastically alters the transport of filaments compared to the ballistic motion of flocks, and we find that clusters of different sizes have different transport behavior. The transport of filaments is measured by the mean-squared displacement (MSD) $\langle \big(\mathbf{r}(t)-\mathbf{r}(0)\big)^2\rangle$. At short times, the MSD for clustered filaments is subdiffusive, and approaches the diffusive regime at late times (Fig.~\ref{fig:msd}). We extracted a diffusion constant from the MSD using the relationship $\langle \big(\mathbf{r}(t)-\mathbf{r}(0)\big)^2\rangle\propto Dt^\alpha$ with $\alpha = 1$ indicating diffusive transport, and superdiffusive and subdiffusive transport being indicated for $\alpha > 1$ and $\alpha < 1$ respectively. The activity of curved filaments leads to circular trajectories and subdiffusive transport at short timescales, and diffusive dynamics at long timescales as filaments are able to escape the confinement of their clusters. 

Achiral filaments with activity have superdiffusive transport due to the ballistic motion of flocks. The short-time MSD for curved active filaments that form clusters is subdiffusive, with a rotational periodicity of the MSD that dampens over time. The long-time behavior of clusters is diffusive. When the radius of curvature is small ($\tilde R \leq 0.5$), the filaments become entangled with a small number of particles~\cite{gravish12} and are kinetically trapped for long times, causing infrequent intermixing between clusters and increasing their stability and lifetime. Comparing the clustering structure and dynamics for homochiral and heterochiral systems, we did not find significant deviations in clustering behavior for the range of parameters examined here (see ESI).

Even at late times, the diffusivity of active filaments is lower than that of passive filaments, indicating that activity leads to an effective freezing of the system dynamics~\cite{reichhardt11}. Flexible filaments with $\tilde R = 0.25$ remain subdiffusive even at long times, likely due to the filaments self-interacting and coiling into spirals (see ESI), which alters the filament MSD compared to clusters of more rigid filaments. The formation of spirals by driven filaments due to self-interactions have been observed in experiments, even for stiff microtubules~\cite{liu11, schaller11}. 

The patterns of clusters in our simulations are similar to those of previous experiments that observed the formation of a lattice of vortices~\cite{schaller11, sumino12}. In our simulations, the size and stability of the vortices vary depending on filament curvature and bending rigidity, and are chirally self-sorted for systems of heterochiral filaments. Unlike previous work with active spinning rotors~\cite{nguyen14, scholz18} or active Brownian particles with Vicsek-type alignment interactions~\cite{liebchen17a, levis19a}, we do not see evidence of macroscopic sorting of clusters into large homochiral domains. Our analysis of the arrangement of the clusters found that the arrangement did not vary significantly from random (see ESI), implying that chiral self-sorting of active curved filaments only occurs at length scales on the order of the filament radius of curvature. This is because the cluster shapes examined here are approximately convex and experience low friction between neighboring domains of opposite chirality. Together with the slow dynamics of filaments with small radius of curvature, macroscopic sorting is unfavorable in these systems.

The results of our Brownian dynamics simulations show that the self-organization of driven semiflexible filaments can undergo a transition between polar flocks and chiral clusters that is tuned by filament curvature and bending rigidity. Our model predicts the collective behavior of future filament gliding experiments for filaments of different persistence lengths that may have intrinsic or induced curvature, and may also have applications for the study of transport behavior for self-assembled curved biopolymers such as FtsZ.

We would like to acknowledge funding for this work from the Soft Materials Research Center under NSF MRSEC Grant No. DMR-1420736, and the National Science Foundation under NSF GRFP Award No. DGE-1144083 and NSF Grant No. DMR-1725065. This work utilized the RMACC Summit supercomputer, which is supported by the National Science Foundation (awards ACI-1532235 and ACI-1532236), the University of Colorado Boulder, and Colorado State University. The Summit supercomputer is a joint effort of the University of Colorado Boulder and Colorado State University.

\bibliographystyle{apsrev4-1}
\bibliography{acf}
\end{document}



\title{\bf{Supplemental Material: Self-organization and chiral self-sorting of active semiflexible filaments with intrinsic curvature}} 
\author{Jeffrey M. Moore}
\author{Matthew A. Glaser} 


\affiliation{Department of Physics, University of Colorado, Boulder, CO 80309}

\author{Meredith D. Betterton}
\affiliation{Department of Physics, University of Colorado, Boulder, CO 80309}
\affiliation{Department of Molecular, Cellular, and Developmental Biology, University of Colorado, Boulder, CO 80309}

\date{\today}

\maketitle

\section{Simulation model}

We model our filaments as discretized wormlike chains~\cite{kratky49} with inextensible segments of length $a$. We have adopted the algorithm by Montesi et. al~\cite{montesi05} for the constrained Brownian dynamics of bead-rod wormlike chains with anisotropic friction. The implementation of the algorithm in our simulations has been covered in previous work~\cite{moore20, moore20a}. What follows here is an overview of the algorithm, as well as the details on our implementation of intrinsic curvature and activity.

Filaments are represented by $N$ sites and $N-1$ segments, with fixed segment length $a$, contour length $L=(N-1)a$, and anisotropic friction, $\zeta_{\bot} = 2\zeta_{\parallel}$. The position of each site $\mathbf{r}_i$ is updated using a midstep algorithm

\begin{equation}\label{posupdate}
  \begin{aligned}
    \mathbf{r}_i^{(1/2)} &= \mathbf{r}_i^{(0)} + \frac{\Delta t}{2}\mathbf{v}_i^{(0)} , \\ 
    \mathbf{r}_i^{(1)} &= \mathbf{r}_i^{(0)} + \Delta t\ \mathbf{v}_i^{(1/2)} ,
  \end{aligned}
\end{equation}
where $\Delta t$ is the time step, $\mathbf{v}_i^{(0)}$ is the initial velocity of site $i$ at the initial position $\mathbf{r}_i^{(0)}$, and $\mathbf{v}_i^{(1/2)}$ is the velocity of site $i$ recalculated at the midstep position $\mathbf{r}_i^{(1/2)}$ with the stochastic forces that were calculated at $\mathbf{r}_i^{(0)}$. The position $\mathbf{r}_i^{(1)}$ is referred to as the fullstep position.

Each site $i$ is assigned an orientation, corresponding to the orientation of the segment attaching it to site $i+1$,
\begin{equation}
  \mathbf{u}_i = \frac{\mathbf{r}_{i+1} - \mathbf{r}_i} { | \mathbf{r}_{i+1} - \mathbf{r}_i | } = \frac{1}{a} (\mathbf{r}_{i+1} - \mathbf{r}_i).
\end{equation}
The orientation of the last site of the filament is set equal to that of its only neighboring segment, so that $\mathbf{u}_N = \mathbf{u}_{N-1}$.

The velocity of each site is
\begin{equation}
  \mathbf{v}_i = \bm{\zeta}^{-1}_{i} \cdot \mathbf{F}^{\text{tot}}_i ,
\end{equation}
where $\bm{\zeta}^{-1}_{i}$ is an anisotropic friction tensor,
\begin{equation}
  \begin{aligned}
    \bm{\zeta}^{-1}_i &= \frac{1}{\zeta_{\parallel}^i} \tilde{\mathbf{u}}_i \otimes \tilde{\mathbf{u}}_i + \frac{1}{\zeta_{\bot}^i} \big(\mathbf{I} - \tilde{\mathbf{u}}_i \otimes \tilde{\mathbf{u}}_i \big) .
  \end{aligned}
\end{equation}
and $\tilde{\mathbf{u}}_i$ is the vector tangent to site $i$, which is the average of the orientations $\mathbf{u}_i$ of its neighboring segments,

\begin{equation}\label{utan}
    \tilde{\mathbf{u}}_i = \frac{(\mathbf{u}_i + \mathbf{u}_{i-1})}{  |\mathbf{u}_i + \mathbf{u}_{i-1} | }
\end{equation}
for $2 \leq i \leq N$, and $\tilde{\mathbf{u}}_1 = \mathbf{u}_1$, $\tilde{\mathbf{u}}_N = \mathbf{u}_{N-1}$ at the chain ends.

In the absence of filament interactions and driving, the total force on site $i$ is the sum
\begin{equation}
  \mathbf{F}^{\text{tot}}_i = \mathbf{F}_i^{\text{bend}} + \mathbf{F}_i^{\text{tension}} + \mathbf{F}_i^{\text{rand}},
\end{equation}
which include bending forces, tension forces, and random forces. The random forces are due to thermal contact with a heat bath at temperature $T$, with the properties $\langle F_\text{rand} \rangle = 0$ and $\langle F^2_\text{rand} \rangle = 2 \zeta k_B T$ to obey the fluctuation dissipation theorem. The random forces are projected onto the chain such that the forces do not conflict with the constraints due to the fixed segment length, and are described in detail by Montesi et al.~\cite{montesi05}.

The diffusivity of a rigid filament is $D = k_B T/\zeta = k_B T/N\zeta_i$, where $\zeta_i$ is the local friction acting on site $i$. The friction depends on the filament aspect ratio $L/\sigma$, where $\sigma$ is the diameter of the chain. In the regime of rigid, infinitely thin rods, the coefficient of friction is given by~\cite{doi88},

\begin{equation}
  \lim_{L/\sigma \to \infty} \zeta_{\bot} = 4 \pi \eta_s L \epsilon.
\end{equation}
where $\eta_s$ is the fluid viscosity. Each site experiences a local friction given by
\begin{equation}
  \zeta_{\bot}^{\text{i}} = 4 \pi \eta_s a \epsilon f(\epsilon).
\end{equation}
where $\epsilon = 1/\ln{(L/\sigma)}$ and
\begin{equation}
  f(\epsilon) = \frac{1+0.64\epsilon}{1-1.15\epsilon}+1.659\epsilon^2.
\end{equation}
is the geometric correction factor for finite aspect ratio filaments.

The bending energy of a discrete wormlike chain for $N \gg 1$ is approximated by
\begin{equation}\label{ubend}
  U_{\text{bend}} = - \frac{\kappa}{a}\sum_{k=2}^{N-1}\mathbf{u}_k\cdot\mathbf{u}_{k-1},
\end{equation}
where $\kappa$ is the bending rigidity, which is related to the persistence length $L_p$ of the wormlike chain as $\kappa = L_pk_BT$. Note that we are adopting the convention that the previous equation is true in all dimensions $d$ of wormlike chains, unlike the convention adopted by Landau and Lifshitz where $\kappa/k_BT = (d-1)L_p/2$~\cite{landau86}. Our convention results in a Kuhn length that depends on dimensionality, $b = (d-1) L_p$.

The bending force is $\mathbf{F}^{\text{bend}}_i=-\partial U_{\text{bend}}/\partial \mathbf{r}_i$. The implementation of the bending forces coincides with metric forces, which come from a metric pseudo-potential that is necessary for the filament conformation to have the expected statistical behavior in the flexible limit, $L_p \ll L$.

\begin{figure*}[tb] \centering
  \includegraphics[width=0.9\textwidth]{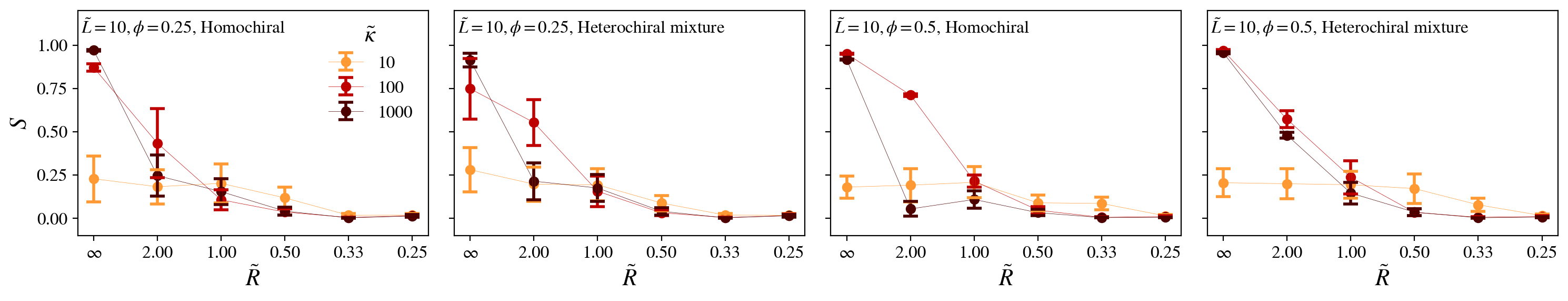}
  \includegraphics[width=0.9\textwidth]{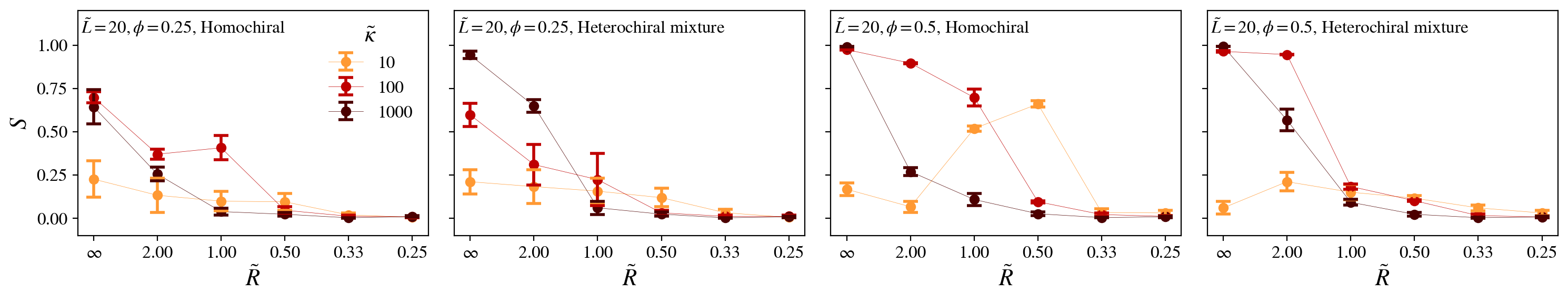}
  \caption{\small Values of the global nematic order $S$ plotted as a function of $\tilde R$ for all simulation parameters. Top row plots are for filaments with aspect ratio $\tilde L=10$ and the bottom row $\tilde L=20$.}
  \label{fig:global_nematic}
\end{figure*}

The bending forces are calculated to include metric forces resulting from a geometric pseudo-potential~\cite{fixman78, montesi05}. The metric pseudo-potential is necessary to observe the proper equilibrium behavior of discrete wormlike chains with low persistence lengths. In the work of Pasquali et. al~\cite{pasquali02}, it was shown that the bending and metric forces together are
\begin{equation}\label{eqn:fbendfmetric}
  \mathbf{F}_i^{\text{bend}}+\mathbf{F}_i^{\text{metric}} = \frac{1}{a}\sum_{k=2}^{N-1}\kappa_k^{\text{eff}}\frac{\partial (\mathbf{u}_k\cdot\mathbf{u}_{k-1})}{\partial \mathbf{r}_i},
\end{equation}
where $\kappa^{\text{eff}}$ is an effective bending rigidity with a conformational dependence,
\begin{equation}\label{keff}
  \kappa_i^{\text{eff}} = \kappa + k_B T a \hat{G}^{-1}_{i-1,i},
\end{equation}
where $\hat G$ is the metric tensor~\cite{pasquali02, montesi05}. The derivative in Eqn.~\ref{eqn:fbendfmetric} can be expanded so that the equation as implemented in our simulation is
\begin{equation}\label{eqn:fbendfmetricx}
  \mathbf{F}_i^{\text{bend}}+\mathbf{F}_i^{\text{metric}} = \frac{1}{a^2}\sum_{k=2}^{N-1}\kappa_k^{\text{eff}}
  \Big((\delta_{i,k+1} - \delta_{i,k})(\mathbf{I}-\mathbf{u}_k\otimes\mathbf{u}_k)\mathbf{u}_{k-1} + (\delta_{i,k} - \delta_{i,k-1})(\mathbf{I}-\mathbf{u}_{k-1}\otimes\mathbf{u}_{k-1})\mathbf{u}_{k}\Big).
\end{equation}

An intrinsic curvature was added to the filament model by modifying the bending potential in Eqn.~\ref{ubend} to have an offset angle $\phi_0$,

\begin{equation}\label{eqn:ubendic}
  U_{\text{bend}} = - \frac{\kappa}{a}\sum_{k=2}^{N-1} \cos{(\theta_{k,k-1} - \phi_0)},
\end{equation}

where $\theta_{k, k-1} = \arccos{(\mathbf{u}_k\cdot\mathbf{u}_{k-1})}$ is the angle between site orientations $k$ and $k-1$, and $\phi_0=ad\phi/ds$ corresponds to the expected angle between two segments of length $a$ with a curvature per unit length $d\phi/ds$.

It can be shown that the term in the sum of Eqn.~\ref{eqn:ubendic} can be rewritten as
\begin{equation}
    \cos{(\theta_{k,k-1} - \phi_0)} = \mathbf{R} \mathbf u_{k} \cdot \mathbf{R}^{-1} \mathbf u_{k-1},
\end{equation}
where $\mathbf R$ is a rotation matrix that rotates the orientation vector $\mathbf u_k$ by an angle $\phi_0/2$,
\begin{equation}
    \mathbf R = 
        \begin{pmatrix}
            \cos(\phi_0/2) & -\sin(\phi_0/2) \\
            \sin(\phi_0/2) & \cos(\phi_0/2)
        \end{pmatrix},
\end{equation}
and its inverse $\mathbf R^{-1}$ rotates the orientation vector $\mathbf u_{k-1}$ by an angle $-\phi_0/2$. The combined bending and metric forces from Eqn.~\ref{eqn:fbendfmetric} with intrinsic curvature are therefore
\begin{equation}
  \mathbf{F}_i^{\text{bend}}+\mathbf{F}_i^{\text{metric}} = \frac{1}{a}\sum_{k=2}^{N-1}\kappa_k^{\text{eff}}\frac{\partial (\mathbf R \mathbf{u}_k\cdot\mathbf R^{-1} \mathbf{u}_{k-1})}{\partial \mathbf{r}_i},
\end{equation}
which can be expanded in the same way as Eqn.~\ref{eqn:fbendfmetricx}.

\begin{figure*}[tb] \centering
  \includegraphics[width=0.9\textwidth]{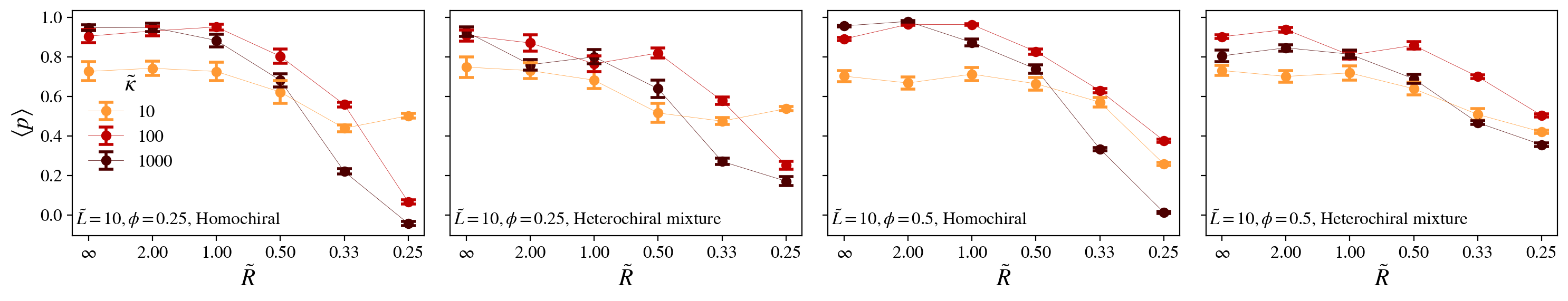}
  \includegraphics[width=0.9\textwidth]{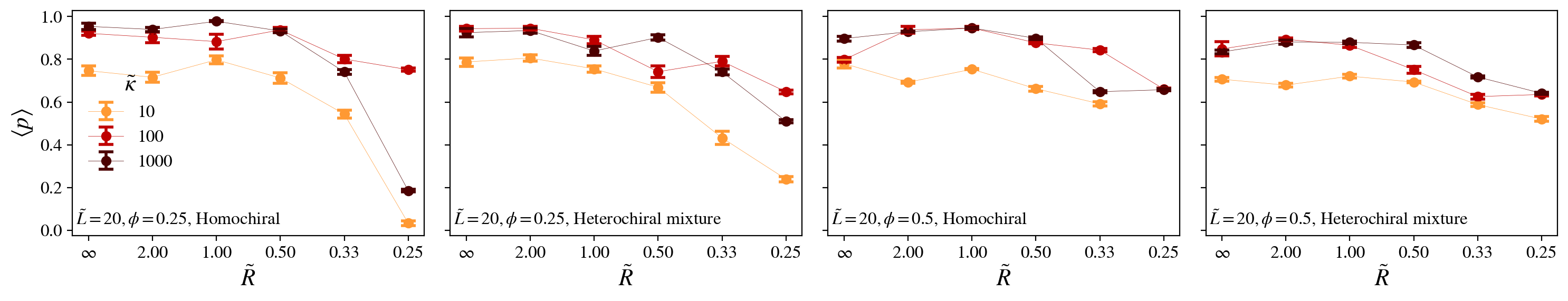}
  \caption{\small Values of the average local polar order $\langle p \rangle$ of simulations plotted as a function of $\tilde R$ for all simulation parameters. Top row plots are for filaments with aspect ratio $\tilde L=10$ and the bottom row $\tilde L=20$.}
  \label{fig:local_polar}
\end{figure*}

Filament driving forces are modeled as a uniform linear force density $f_{\text{dr}}$ that is directed along the local filament segment orientations,
\begin{equation}
  \mathbf{F}_{\text{dr}} = f_{\text{dr}}\mathbf{u}_i.
\end{equation}
The assumptions of this model match observations of experiments with gliding filaments driven by a lattice of motor proteins, which found that that filament velocities were constant, despite the persistent binding and unbinding of motors~\cite{liu11}.

\section{Model implementation}

Simulation software for the filament model is written in C++ and the source code is publicly available online ~\cite{moore20}. The software is also available as a pre-installed binary on Singularity and Docker images. The simulations were run on the Summit computing cluster~\cite{anderson17} and parallelized using OpenMP.

\begin{figure*}[tb] \centering
  \includegraphics[width=0.9\textwidth]{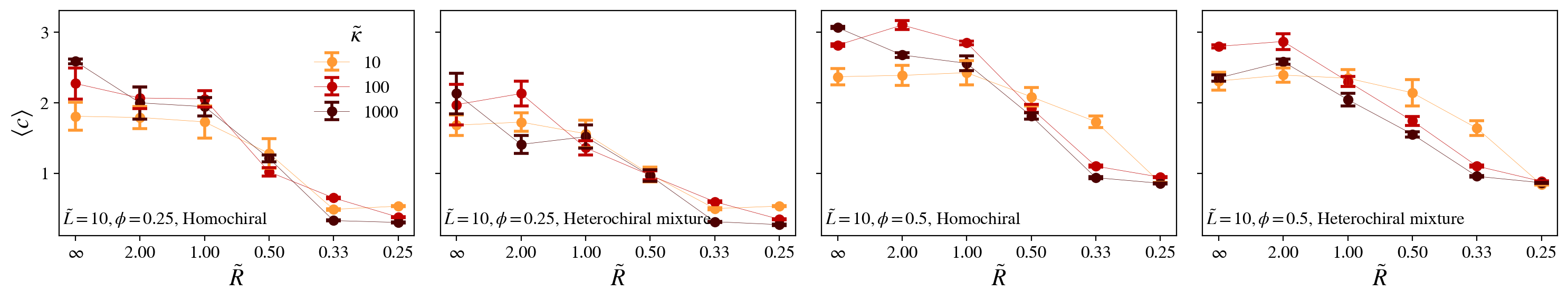}
  \includegraphics[width=0.9\textwidth]{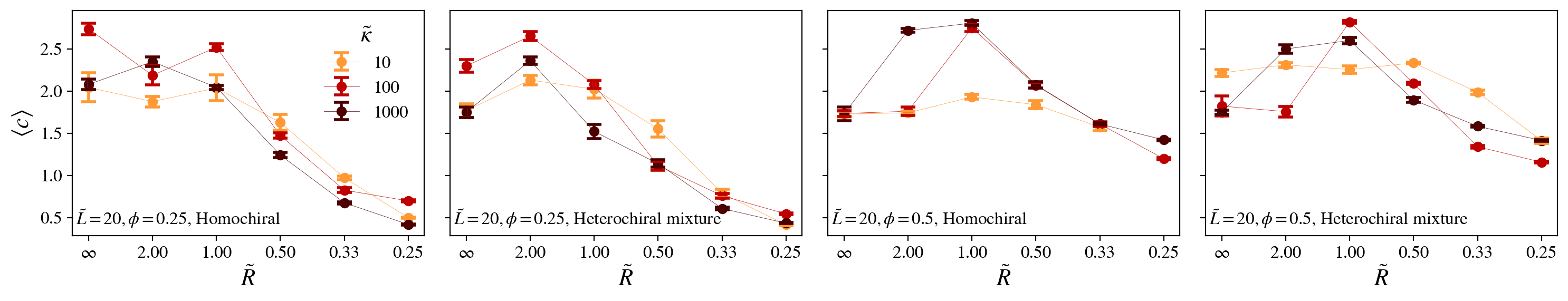}
  \caption{\small Values of the average contact number $\langle c \rangle$ plotted as a function of $\tilde R$ for all simulation parameters. Top row plots are for filaments with aspect ratio $\tilde L=10$ and the bottom row $\tilde L=20$.}
  \label{fig:contact_number}
\end{figure*}

\section{Simulation parameters}

Important parameters of our simulation are the filament contour length $L$, diameter $\sigma$, bending rigidity $\kappa$, driving force per unit length $f_{dr}$, filament radius of curvature $R$, simulation box diameter $L_{sys}$, and filament density $\phi$. Our simulations have filament aspect ratios $\tilde L = L/\sigma = 10$ and $20$, and the system size is $~\tilde L_{sys} = L_{sys}/L = 10$. In our dimensionless reduced units, $\sigma$, $k_B T$, and $D$ are set to be unity, where $D$ is the diffusion coefficient for a sphere of diameter $\sigma$, such that the viscosity is $1/3\pi$. The driving force in reduced units is $f_{dr} = 15$, such that the P\'eclet number is $\text{Pe} = f_{dr}L^2/k_B T \approx 5\times 10^4$, which was chosen to avoid issues arising from the effects of filament softening due to tangential driving~\cite{isele-holder15, anand18, gupta19, peterson20, moore20a}.

The dimensionless parameters used in our analysis are $\tilde{\kappa} = \kappa/Lk_BT = L_p/L$, where $L_p$ is the filament persistence length, the filament radius of curvature $\tilde{R} = R/L$, and filament density in terms of the particle packing fraction $\phi = A_\text{fil}/A_\text{sys}$, where $A_{\text{sys}}$ is the area of the 2D periodic simulation space and $A_{\text{fil}}=N(L\sigma + \pi\sigma^2)$ is the area occupied by $N$ 2D spherocylindrical filaments.

We used a dynamic timestep in the half-step integration algorithm, with a maximum timestep $\Delta t = 2.5\times 10^{-5} \tau$, where $\tau$ is the average time for a sphere of diameter $\sigma$ to diffuse its own diameter. If ever forces between any two particles ever exceed a preset threshold of $10^6$ reduced force units, all particles are returned to the previous full-step positions, the timestep is reduced by a factor of 2, and forces are recalculated. The time resolution of filament positions for the purposes of analysis are fixed to be $\Delta t_{\text{max}}$. The active timescale used in our analysis is the time required for a straight filament to glide its own length $\tau_A = l / v_{dr} = 1 / \zeta_\parallel f_{dr}$, which is $0.66 \tau$ for $\text{Pe} = 5 \times 10^4$.

Filaments in the simulation were initialized by randomly inserting straight filaments parallel to one axis of the simulation box in a nematic arrangement, allowing the filaments to relax and diffuse without activity for $100 \tau$ before introducing driving forces. Simulations terminated once they were determined to have reached a steady state, when order parameters appeared to converge to constant values. 

\section{Flocking analysis}
Long-range structural order in our simulations filaments is captured by the nematic order parameter of the system $S$, which is the largest eigenvalue of the 2D nematic order tensor 
\begin{equation}
  \mathbf{Q}=\frac{1}{N}\sum_{i=1}^{N} (2\mathbf{u}_i\otimes\mathbf{u}_i - \mathbf{I}),
\end{equation}
where $\mathbf{I}$ is the unit tensor. High nematic order indicates that flocks have aggregated into giant flocks, which tend to dominate the overall system structure. Nematic order is present for rigid filaments and large radius of curvature (Fig.~\ref{fig:global_nematic}).

Although curvature and flexibility inhibit long-range order, polar flocks are present at all but the highest curvatures examined here, $\tilde R = 0.25$. Following previous work, flocking behavior was identified by measuring the filament contact number $c_i \sum_{i\neq j}= e^{-(r_{ij}/\sigma)^2}$ and the local polar order parameter $p_i = \sum_{i \neq j} \mathbf{u}_i \cdot \mathbf{u}_j e^{-(r_{ij}/\sigma)^2}/c_i$, with sums ranging over all filament segments, excluding intrafilament segments. The time and ensemble-average of the local polar order all simulation parameters is plotted in Fig.~\ref{fig:local_polar}.

Systems with polar-ordered collective motion exhibit giant number fluctuations (GNF)~\cite{gregoire04, chate08, ginelli10, ginelli16}. Number fluctuations are derived from the mean $\langle N \rangle$ and standard deviation $\Delta N$ of the particle number within a subregion of the system. Varying the size of the region leads to a power-law scaling behavior $\Delta N \propto \langle N \rangle^\alpha$. For equilibrium systems, number fluctuations scale with the exponent $\alpha = 1/2$, whereas systems with GNF exhibit scaling with $\alpha > 0.5$, with the Vicsek model having $\alpha \approx 0.8$~\cite{chate08, ginelli10}.

\begin{figure*}[tb] \centering
  \includegraphics[width=0.9\textwidth]{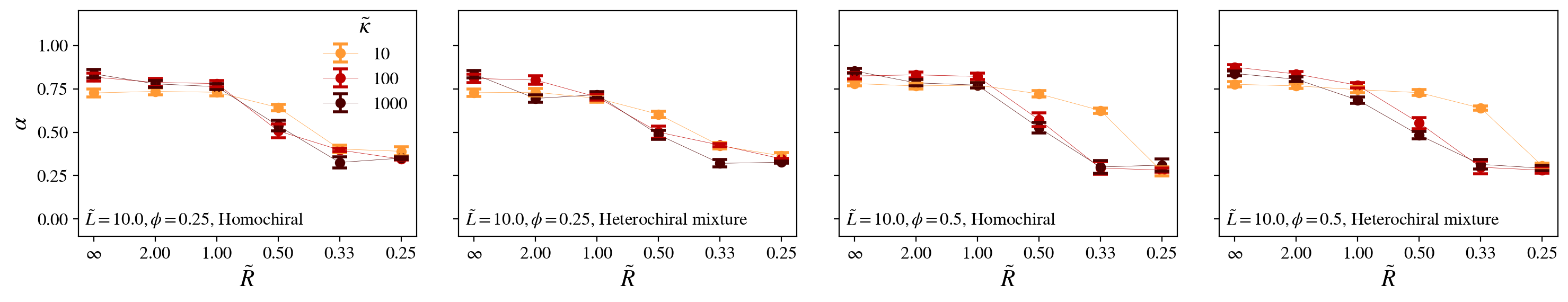}
  \includegraphics[width=0.9\textwidth]{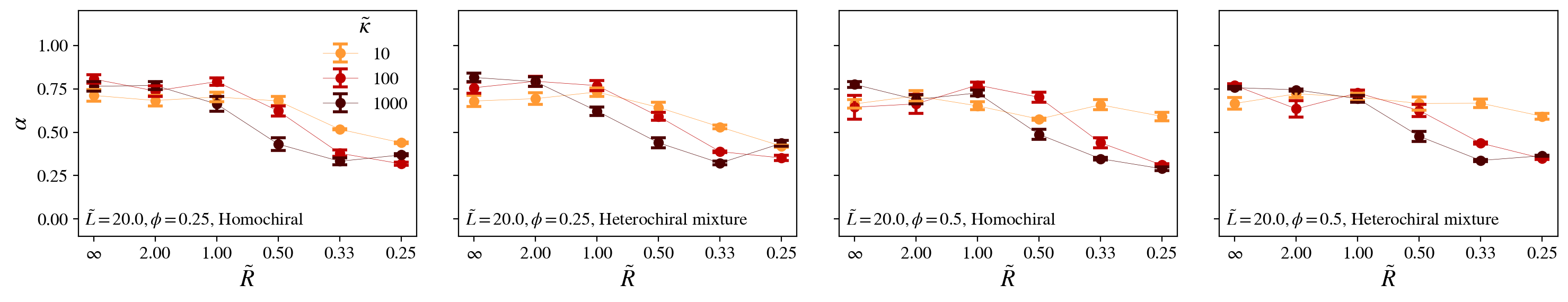}
  \caption{\small Values of the exponential scaling $\alpha$ for the number fluctuations in the system $\Delta N\propto\langle N \rangle^\alpha$ plotted as a function of $\tilde R$ for all simulation parameters. Top row plots are for filaments with aspect ratio $\tilde L=10$ and the bottom row $\tilde L=20$.}
  \label{fig:gnf}
\end{figure*}

The number fluctuation scaling for all simulations is plotted in Fig.~\ref{fig:gnf}. In the flocking regime with straight filaments, filaments exhibit GNF with $\alpha \approx 0.8$. The number fluctuations decrease with increasing filament curvature, and in some cases enters a regime with $\alpha < 0.5$ indicating subdiffusive behavior, causing small density fluctuations at short timescales.

\begin{figure*}[tb] \centering
  \includegraphics[width=0.9\textwidth]{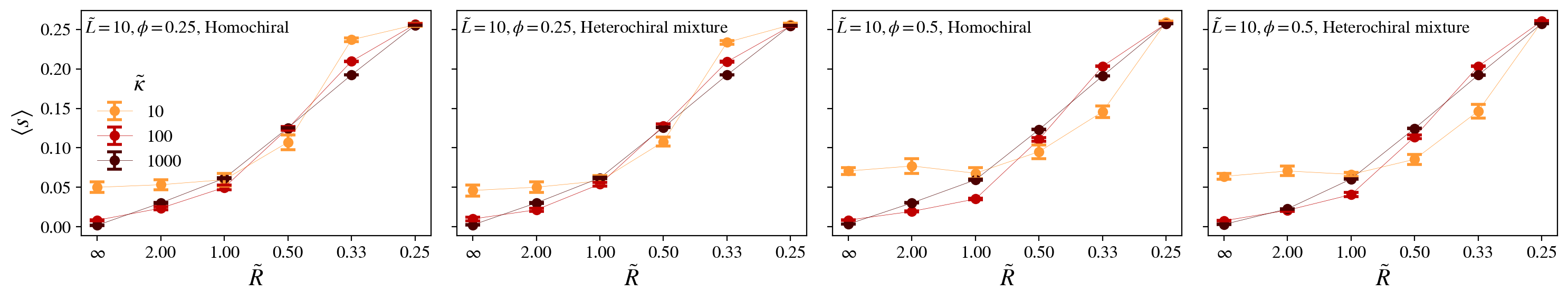}
  \includegraphics[width=0.9\textwidth]{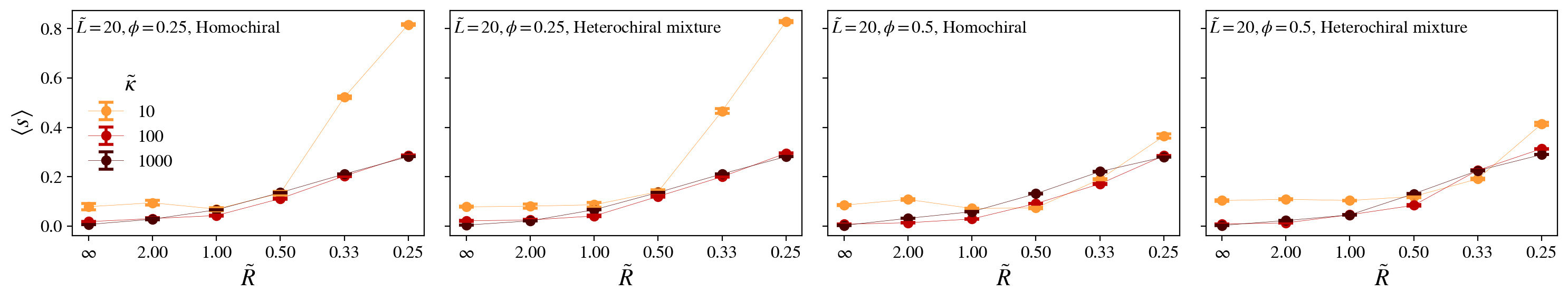}
  \caption{\small Values of the average spiral number $\langle s \rangle$ plotted as a function of $\tilde R$ for all simulation parameters. Top row plots are for filaments with aspect ratio $\tilde L=10$ and the bottom row $\tilde L=20$.}
  \label{fig:global_spiral}
\end{figure*}

For a driven flexible filament, there is a chance for the filament to self-interact and wrap upon itself, winding into a spiral-like structure. We have previously measured the spiral-similarity of bent filaments using a spiral number $\langle s \rangle$~\cite{moore20a}. The spiral number for an individual filament is calculated by measuring the angle $\theta_i$ swept by traversing its contour length from tail to head originating from the center of curvature of the filament, $s_i = \frac{1}{2\pi}\theta_i$. A straight filament will have a spiral number $s_i = 0$, a filament bent into a perfect circle has $s_i=1$, and filaments that form tightly-wound spirals may have a spiral number $s_i>1$.

The average spiral number for curved filaments at equilibrium will reflect the radius of curvature of the filament. However, effects due to interactions, driving, and flexibility will modify the overall spiral number. We find that flexible filaments with aspect ratio $\tilde L=10$ have higher spiral numbers than more rigid filaments at large $\tilde R$. However, we surprisingly find that flexible filaments with intermediate $\tilde R$ have a slightly smaller spiral number than the most rigid filaments in our simulations (Fig.~\ref{fig:global_spiral}). This is likely due to the flexible filaments forming heterochiral flocks, while rigid filaments only form homochiral clusters. For filaments with aspect ratio $\tilde L=20$, filaments have a much higher spiral number compared to other rigidities when filaments have small radius of curvature, $\tilde R=0.25$, indicating the formation of tightly-wound and dynamically frozen spirals. The formation of these structures is likely the cause of the subdiffusive behavior for flexible filaments even at long times.

\section{Mean-squared displacement}
\begin{figure*}[tb] \centering
  \includegraphics[width=\textwidth]{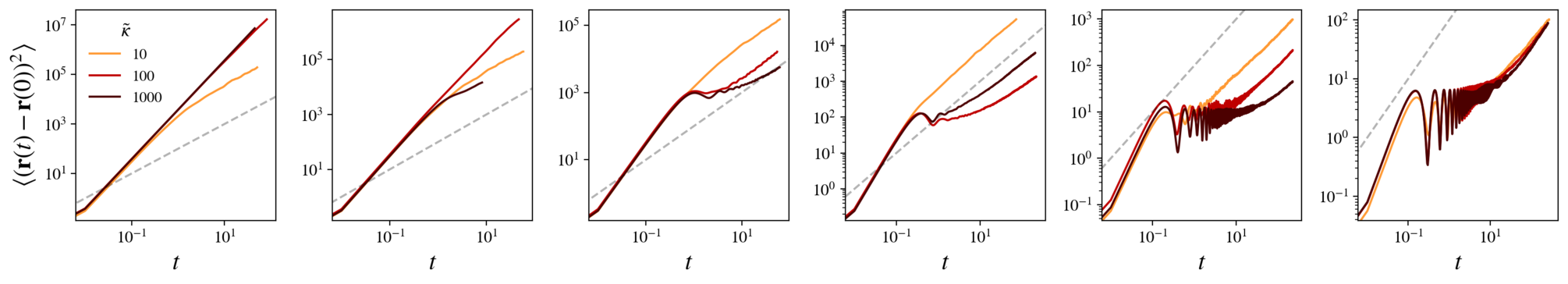}
  \caption{\small Mean-squared displacements (MSDs) plotted for $\tilde L=10$ for heterochiral systems with $\phi=0.25$, showing the short- and long-time scaling behavior. The effective diffusion coefficent $D_{\text{active}}$ was found by fitting the final $25\tau_A$ of the MSD.}
  \label{fig:msd}
\end{figure*}

The mean-squared displacements (MSD) of filaments for both inactive and active filaments were calculated using the equation $\langle \big(\mathbf{r}(t) - \mathbf{r}(t_0)\big)^2\rangle$, where the brackets $\langle ... \rangle$ denote an average over the ensemble of filaments and time averages for different values of $t_0$ separated by a minimum of $10\tau_A$. Example MSDs for $\tilde L=10$ are plotted in Fig.~\ref{fig:msd} on a log-log scale, with the gray dashed line denoting linear time-scaling behavior.

The effective diffusion coefficient for active filaments $D_{\text{active}}$ was calculated using the final $25\tau_A$ interval of the MSD to limit analysis to long-time transport behavior of filaments, assuming a linear time scaling. To determine whether the long-time behavior was diffusive, we calculated the power-law scaling of the effective diffusion coefficient with respect to time, $\langle \big(\mathbf{r}(t) - \mathbf{r}(t_0)\big)^2\rangle \propto D t^\alpha$ by measuring the slope of the log-log transform of the MSD using a weighted least squares linear regression model, with weights derived from the standard error of the mean for values of the MSD.

\section{Identification of filament clusters}
To quantify the dynamics and structure of the filament clusters, filaments were clustered by their centers of curvature $\mathbf{r}_c(t)$, determined from the filaments' instantaneous radius of curvature $R(t)$ averaged over the contour length of the filament. Cluster positions are defined to be the average of their constituent filament centers of curvature, $\mathbf{r}_C(t) = \frac{1}{n}\sum_{i}^n \mathbf{r}_c^{(i)}(t)$, and the cluster radii are defined to be the average of the constituent filament curvature radii $R_c(t) = \frac{1}{n}\sum_i^n R_i(t)$.

Unclustered filaments can join an existing cluster when $|\mathbf{r}_c^{(i)}(t) - \mathbf{r}_C(t)| \leq R_c(t)$ for a time interval of $\tau_A$. Two previously unclustered filaments can form a new cluster if their centers of curvature are bounded by the average of their curvature radii, $|\mathbf{r}_c^{(i)}(t) - \mathbf{r}_c^{(j)}(t)| < (R_i(t) + R_j(t))/2$ for a minimum time interval of $\tau_A$. Filaments can leave a cluster if their center of curvature leaves the bounded space defined by the curvature position and curvature radius for an interval of $0.25 \tau_A$, or if ever the filament center of curvature is no longer oriented in the direction of the cluster position, $(\mathbf{r}_c(t) - \mathbf{r}_i(t)) \cdot (\mathbf{r}_C(t) - \mathbf{r}_i(t)) < 0$. A cluster is annihilated if ever the number of filaments in the cluster is less than 2.

\begin{figure*}[p!] \centering
  \includegraphics[width=\textwidth]{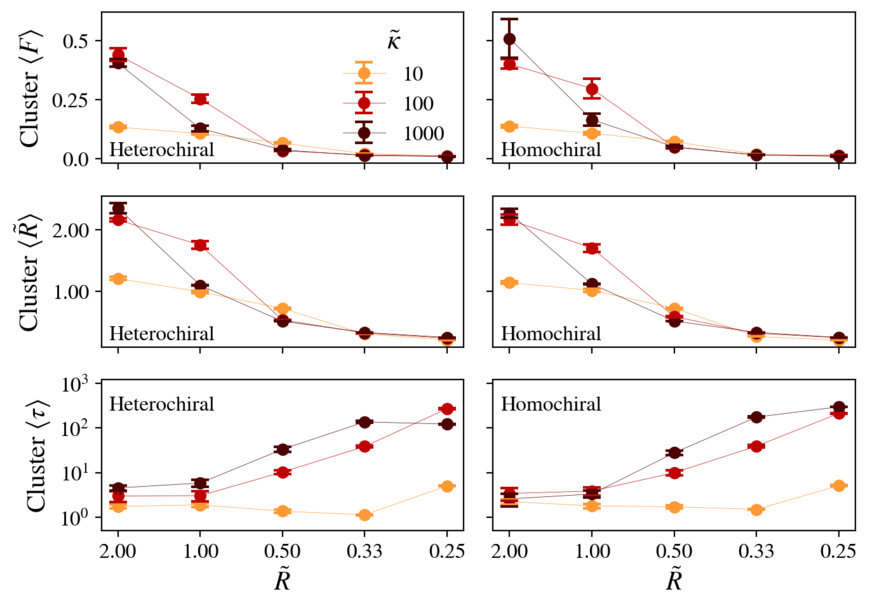}
  \caption{\small Top row: the fraction of simulation filaments in an average cluster $\langle F \rangle$ for homochiral and heterochiral systems. Center row: average cluster radius $\langle \tilde R\rangle$ plotted for homochiral and heterochiral systems. Bottom row: Average cluster lifetime $\langle \tau \rangle$ for homochiral and heterochiral systems plotted with respect to radius of curvature. Lifetimes are expressed in units of $\tau_A$.}
  \label{fig:cluster_comparison}
\end{figure*}

Fig.~\ref{fig:cluster_comparison} compares the average cluster radius $\langle \tilde R \rangle$ (normalized by the filament length) for homochiral and heterochiral systems. At large filament curvature radii, slightly larger clusters appear to be possible, which are perhaps limited by finite size effects. However, there does not appear to be a significantly different cluster scaling between heterochiral and homochiral systems.

\section{Randomness of filament clusters}
We assessed whether the clusters of filaments with small radius of curvature $\tilde R \leq 0.5$ sorted macroscopically into larger domains of homochiral clusters by measuring the mixing between left-handed (LH) and right-handed (RH) clusters. Upon identifying the cluster positions, described above, we constructed an adjacency matrix $X$ representing a graph with cluster centers as vertices and edges joining the cluster nearest neighbors. In a well-mixed (random) system, the handedness of nearest neighbors for any one vertex should be $\pm1$ with equal probabilities. This would imply that each vertex of $X$ would have adjacent neighbors with a net handedness $\Sigma = \sum_i^{\text{adj}}\chi_i$ that should be zero on average but with normal variance from a randomly distributed network. In a system with sorted domains, the nonrandom distribution of handedness among clusters would give rise to a bimodal distribution of $\Sigma$, and a nonrandom lattice with approximately alternating handedness would be unimodal with zero mean and very small variance.

\begin{figure*}[htb] \centering
  \includegraphics[width=\textwidth]{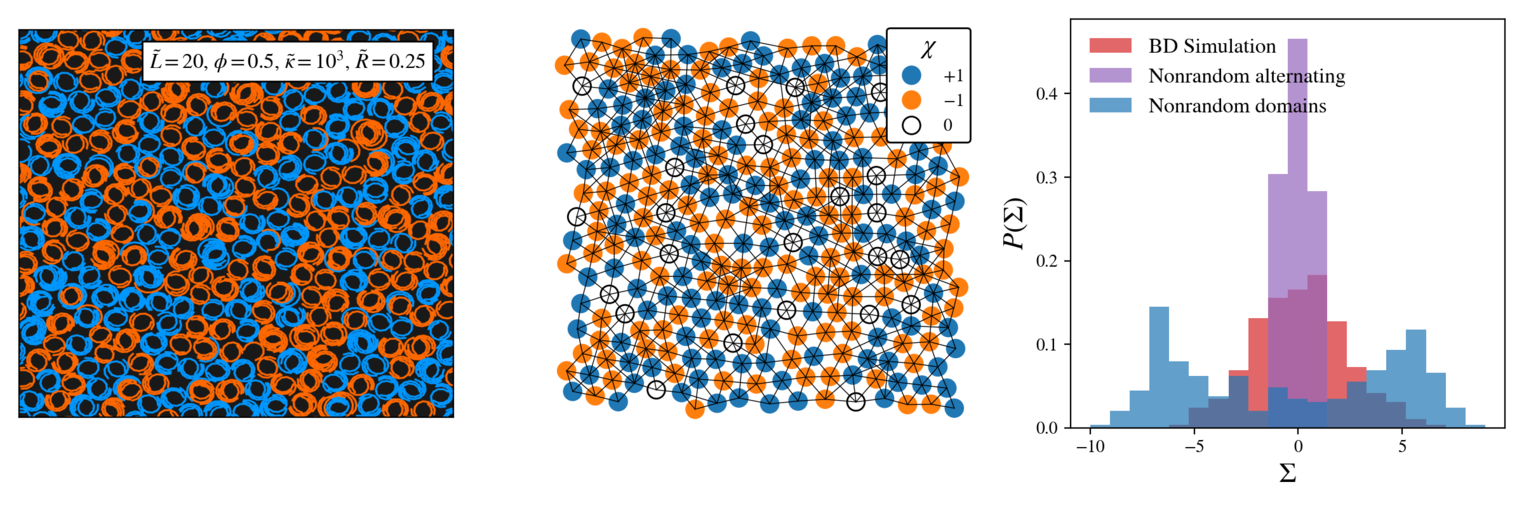}
  \caption{\small Left: simulation image for a heterochiral mixture of homochiral clusters. Center: The associated adjacency graph joining the centers of clusters to their nearest neighbors, with with clusters of mixed handedness being assigned a handedness of zero. The graph does not show edges between vertices across periodic boundaries, although these edges were present in our analysis. Right: the distribution of the sum of neighbor handedness $\Sigma$ for each vertex, plotted for the associated simulation to the left (shown in red), contrasted with distributions for the same adjacency graph with hypothetical nonrandom handedness distributions.}
  \label{fig:entropy}
\end{figure*}

In Fig.~\ref{fig:entropy} the distribution of $\Sigma$ (right) associated with the simulation image (left) is shown in red. The distribution appears normal with zero mean, and is contrasted with distributions of $\Sigma$ for nonrandom distributions of handedness. The adjacency graph associated with the image is plotted in the center. There does not appear to be any sign of macroscopic sorting in the distribution, so we must conclude that the distribution of handedness among the clusters is random. This approach was repeated for different simulation parameters, without any indication of nonrandomness.

\section{Diagrams of simulation images}

\begin{figure*}[p!] \centering
  \includegraphics[width=\textwidth]{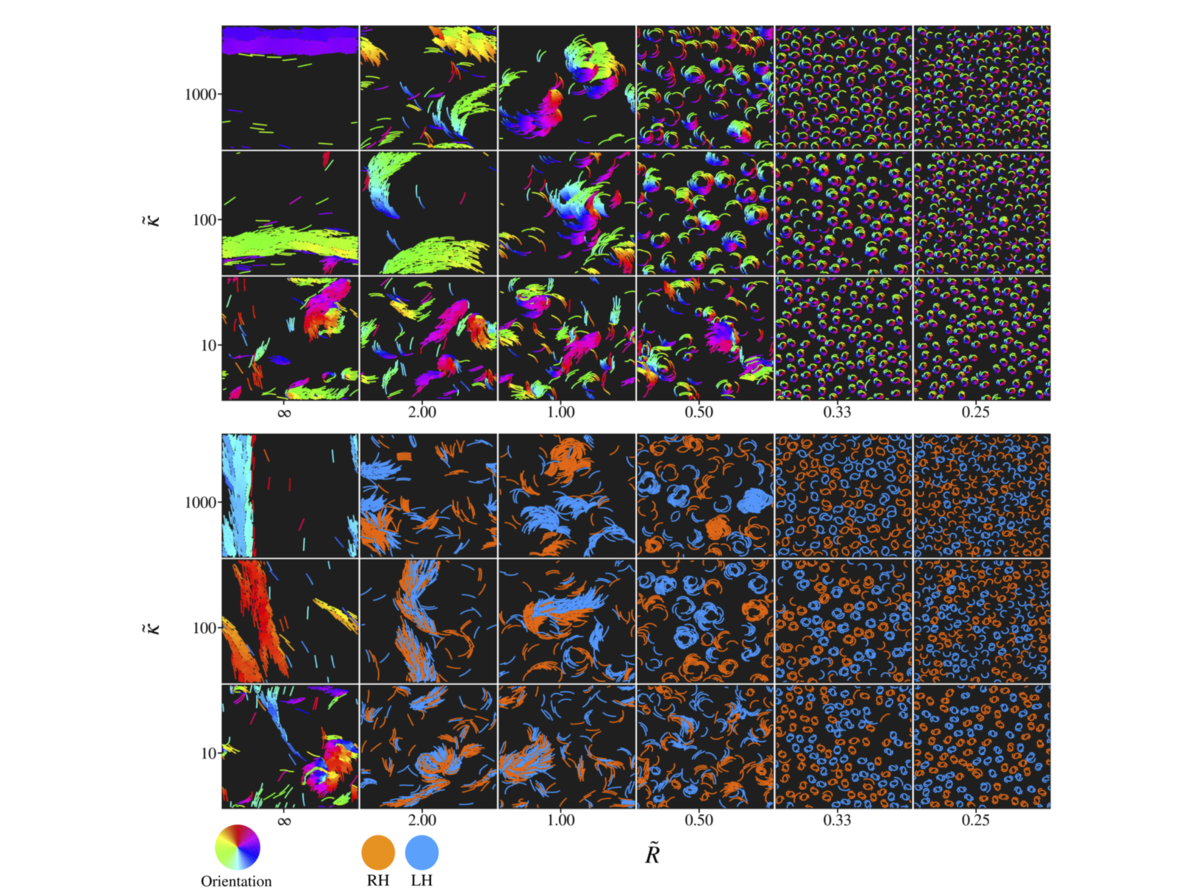}
  \caption{\small Simulation images for filaments with $\tilde L=10$ and packing fraction $\phi=0.25$ for homochiral (top) and heterochiral (bottom) systems.}
  \label{fig:first}
\end{figure*}

\begin{figure*}[htb] \centering
  \includegraphics[width=\textwidth]{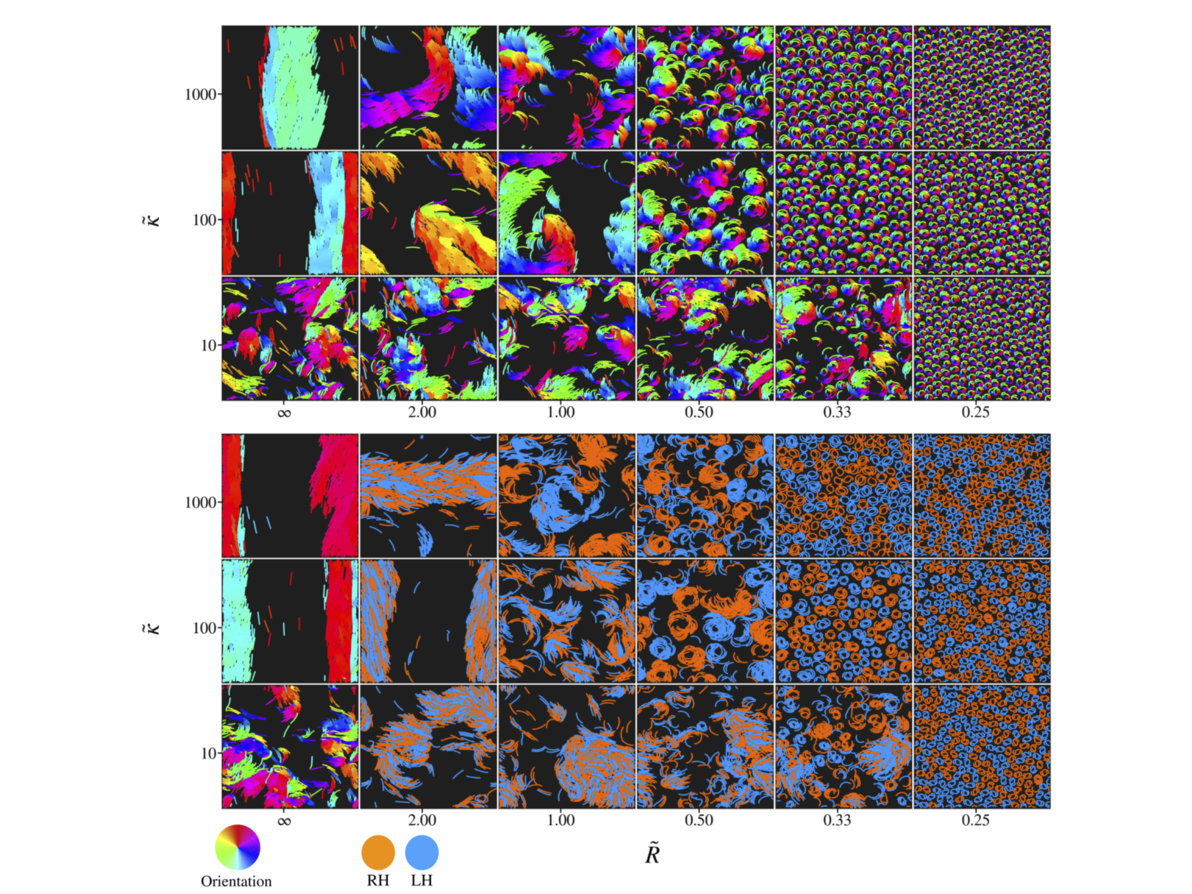}
  \caption{\small Simulation images for filaments with $\tilde L=10$ and packing fraction $\phi=0.50$ for homochiral (top) and heterochiral (bottom) systems.}
  \label{fig:second}
\end{figure*}

\begin{figure*}[p!] \centering
  \includegraphics[width=\textwidth]{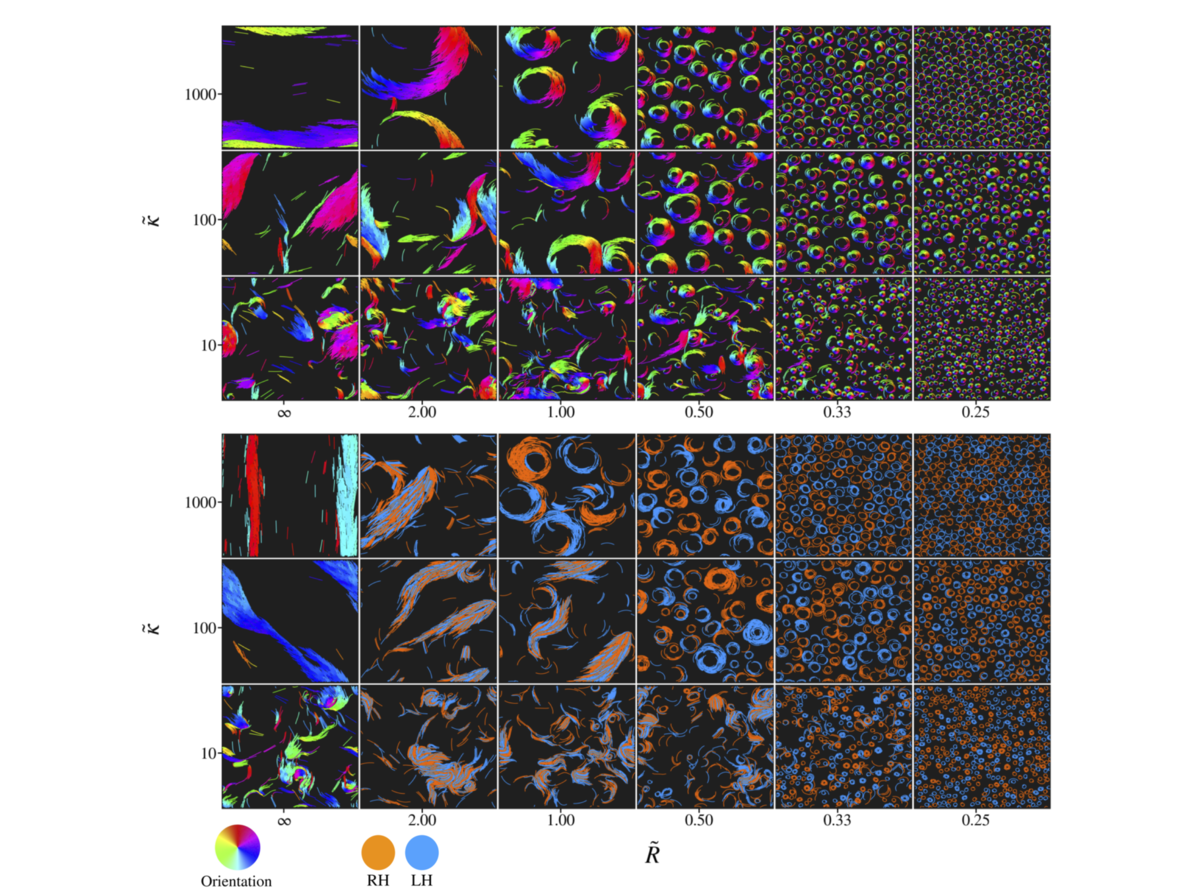}
  \caption{\small Simulation images for filaments with $\tilde L=20$ and packing fraction $\phi=0.25$ for homochiral (top) and heterochiral (bottom) systems.}
  \label{fig:third}
\end{figure*}

\begin{figure*}[htb] \centering
  \includegraphics[width=\textwidth]{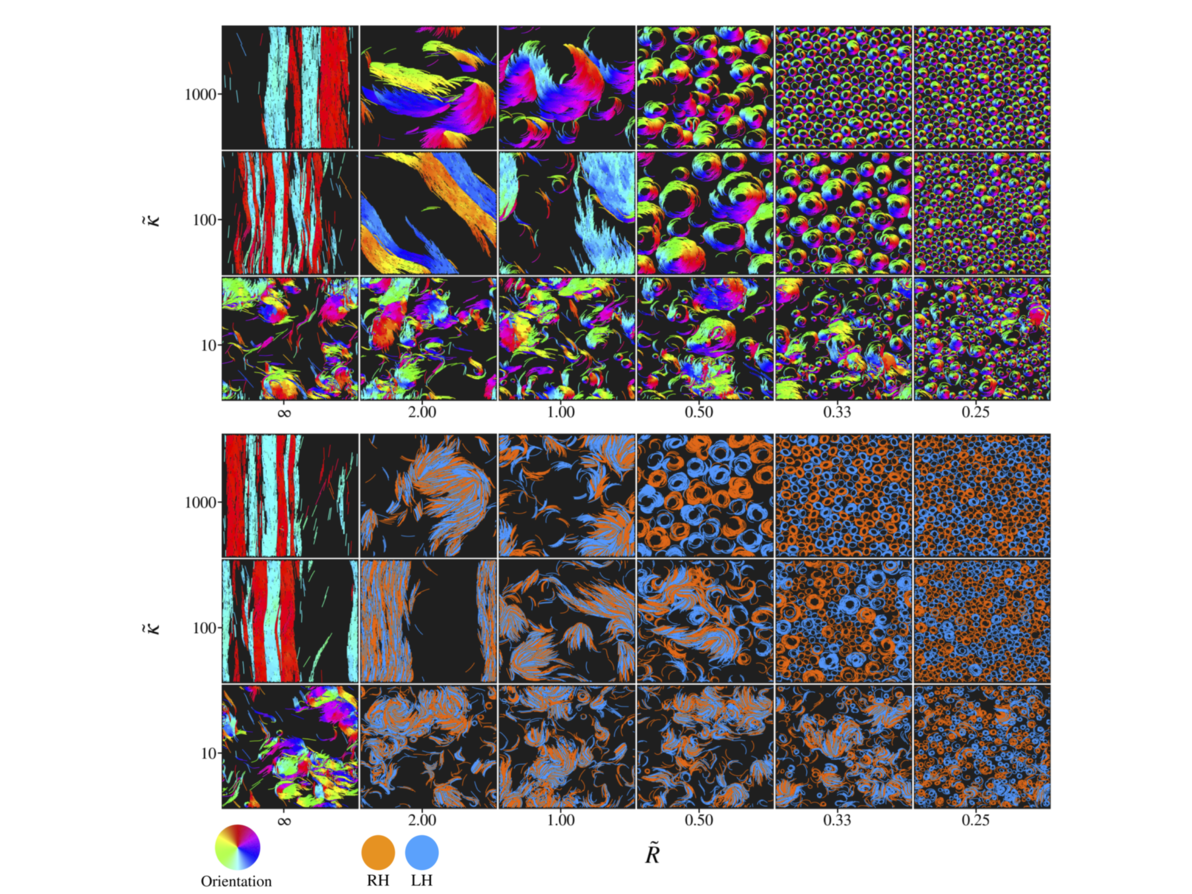}
  \caption{\small Simulation images for filaments with $\tilde L=20$ and packing fraction $\phi=0.50$ for homochiral (top) and heterochiral (bottom) systems.}
  \label{fig:last}
\end{figure*}

Figs.~\ref{fig:first}--\ref{fig:last} are diagrams of simulation images for the collective behavior displayed for varying $\tilde R$ and $\tilde \kappa$ for homochiral and heterochiral filaments at filament densities $\phi=0.25$ and $0.5$ and filament aspect ratios $\tilde L=10$ and $20$. 

Noteably, Fig.~\ref{fig:first} and Fig.~\ref{fig:last} have images of filaments with $\tilde R=2$ that highlight the breakdown of long-range polar order for rigid filaments ($\tilde \kappa=1000$) due to the filament packing, as mentioned in the main text.

\bibliographystyle{unsrt} \bibliography{acf} 

 \clearpage

\clearpage

 \clearpage

\clearpage

 \clearpage

\clearpage